# Evaluation of Controllers and Development of a new in-house Controller for the Teledyne HxRG Focal Plane Array for the IRSIS satellite payload

M. B. Naik[1], D. K. Ojha[1], S. K. Ghosh[1], P. Manoj[1], J. P. Ninan[1], S. Ghosh[1], S. L. A. D'Costa[1], S. S. Poojary[1], S. B. Bhagat[1], P. R. Sandimani[1], H. Shah[1], R. B. Jadhav[1], S. M. Gharat[1], G. S. Meshram[1] and B. G. Bagade[1]

[1] *Department of Astronomy & Astrophysics, Tata Institute of Fundamental Research, Mumbai, 400005, India, mbnaik@tifr.res.in*

The Infrared Astronomy Group (Department of Astronomy and Astrophysics) at Tata Institute of Fundamental Research (TIFR) is presently developing controllers for the Teledyne HxRG Focal Plane Arrays (FPAs) to be used on board the Infrared Spectroscopic Imaging Survey (IRSIS) satellite payload. In this manuscript we discuss the results of our tests with different FPA controllers like the Astronomical Research Cameras (ARC) controller, Teledyne's SIDECAR ASIC as well as our new in-house designed Array controller. As part of the development phase of the IRSIS instrument, which is an optical fibre based Integral Field Unit (IFU) Near-Infrared (NIR) Spectrometer, a laboratory model with limited NIR bandwidth was built which consisted of various subsystems like a Ritchey-Chretien (RC) 30 cm telescope, optical fibre IFU, spectrometer optics, and the Teledyne H2RG detector module. We discuss the various developments during the building and testing of the IRSIS laboratory model and the technical aspects of the prototype in-house H2RG controller.

## 1. Introduction

The Infrared Astronomy Group (Department of Astronomy and Astrophysics) at Tata Institute of Fundamental Research (TIFR) has been using the Teledyne H1RG and H2RG focal plane arrays (FPAs) in ground-based near-infrared (NIR) Imagers and Spectrometers installed on 2- and 4-meter class telescopes in India. Proposed NIR spectrometers like the ground-based Multi-Object Infrared Spectrometer and the space-based Infrared Spectroscopic Imaging Survey (IRSIS) satellite payload are designed with the Teledyne NIR HgCdTe (Mercury Cadmium Telluride) FPAs. The IRSIS payload is being planned as a Small Satellite Payload to be launched into a low earth (~800 km) sun synchronous orbit by a Polar Satellite Launch Vehicle of the Indian Space Research Organization (Ghosh, 2010). The instrument will consist of a 30 cm Ritchey-Chretien (RC) telescope with an optical fibre based integral field unit (IFU) feeding into two spectrometer channels: Shortwave (1.7 to 3.4 μm) and Longwave (3.2 to 6.4 μm) with a moderate spectral resolution ($R = \lambda/\Delta\lambda \sim 100$) and a sensitivity of 14 magnitude in the K band (2.2 μm). The IRSIS will function as a survey instrument with simultaneous spectroscopy in the two bands with a large field of view of 14 arc min x 14 arc min. The IRSIS will cover about 50 % of the sky in a period of 2 years and is expected to lead to many significant astrophysical results due to the unexplored nature of the spectral range covered. The primary science goals include: (1) detection of several spectral features due to the interstellar dust and gas components in our Galaxy; (2) first ever infrared classification of stellar populations in our Galaxy; (3) Complete census of low mass objects in the Solar neighbourhood (~30 pc); (4) Unexpected discoveries / surprises from this new uniform survey covering a major fraction of the full sky. A laboratory model limited to a part of the shortwave channel (1.7 to 2.5 μm), was built to demonstrate the concept and the capability to build such an instrument.

The HxRG FPAs (Beletic *et al.*, 2008) are developed by Teledyne Imaging Sensors (TIS), USA[a]. HxRG is an acronym for HgCdTe Astronomy Wide Area Infrared Imager, with Reference pixels and Guide mode. The HgCdTe material has a band gap energy that can be tuned from 0.1 to 0.5 eV by varying the relative fraction of cadmium and is responsive from about 1 to some tens of μm. Fig. 1 shows the functional grade H2RG FPA in our test dewar at TIFR. The FPA is a hybrid device wherein the HgCdTe pixel is connected to the Silicon Readout Integrated Circuits (ROIC) via Indium bumps. The ROIC is an integrated circuit which contains all the transfer electronic circuitry wherein the electrons generated by light falling on the HgCdTe layer are transferred to the underlying ROIC for readout through a charge to voltage amplifier. The bare ROIC is also available separately and can be used to test the control and data handling electronics at room temperature before the actual FPA is connected. Since the ROIC is Silicon based, it is sensitive to visible light and so visible light images can be taken using an ROIC at low efficiency. The H1RG can be readout from either 1, 2 or 16 channels simultaneously whereas the H2RG can be readout from 1, 4 or 32 channels simultaneously. Additional details on H2RG ROIC and FPA are given in Appendix A.

---

[a] www.teledyne-si.com/business-units/teledyne-imaging-sensors

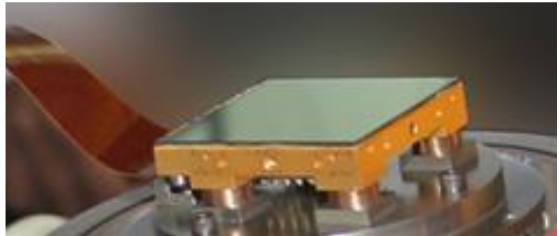

Fig. 1. H2RG FPA with a flex cable in the test dewar

In this paper, Section 2 provides a brief overview of the IRSIS instrument and the laboratory model. Section 3 describes the available FPA controllers. Sections 4 and 5 describe calibration of H2RG ROIC and H2RG FPA respectively. Section 6 discusses the new in-house FPA controller design along with the test results and Section 7 summarizes the results and the work remaining to be done in future.

**2. The IRSIS instrument and the Laboratory Model:**

A block diagram of the IRSIS instrument is shown in Fig. 2. The telescope consists of a 30 cm diameter primary mirror in RC configuration. At the focal plane of this telescope, two fibre based IFU of 1000 fibres each, coupled through microlenses, will reimage the focal plane into 5 slits for each channel. Each slit will consist of 200 fibres and will serve as input to the spectrometer for that particular channel. The Shortwave channel will be sensitive in the 1.7 to 3.4 μm band and the Longwave channel will be sensitive in the 3.2 to 6.4 μm band. The two spectrometers will consist of the collimator optics, the dispersive element which will be transmission grating and the camera optics to focus the spectra onto the NIR FPA which will be the Teledyne 1024 x 1024 pixel H1RG. The entire telescope and spectrometer optics will be cooled to 100 Kelvin and the shortwave FPA will be cooled to 80 Kelvin and the longwave FPA will be cooled to 50 Kelvin using passive cooling methods. Fig. 3 shows the optical design layout by the Zemax ray tracing software (ZEMAX, LLC) for one spectrometer channel. To enable IRSIS to achieve the desired operating temperatures entirely by passive cooling, the satellite will be launched into a 6 am to 6 pm or Dawn to Dusk Sun Synchronous orbit shielded from the Sun's radiation using a V-groove structure (Cornut *et al.*, 2000; Bolton, 2017; Hasebe *et al.*, 2018) as shown in Fig. 4.

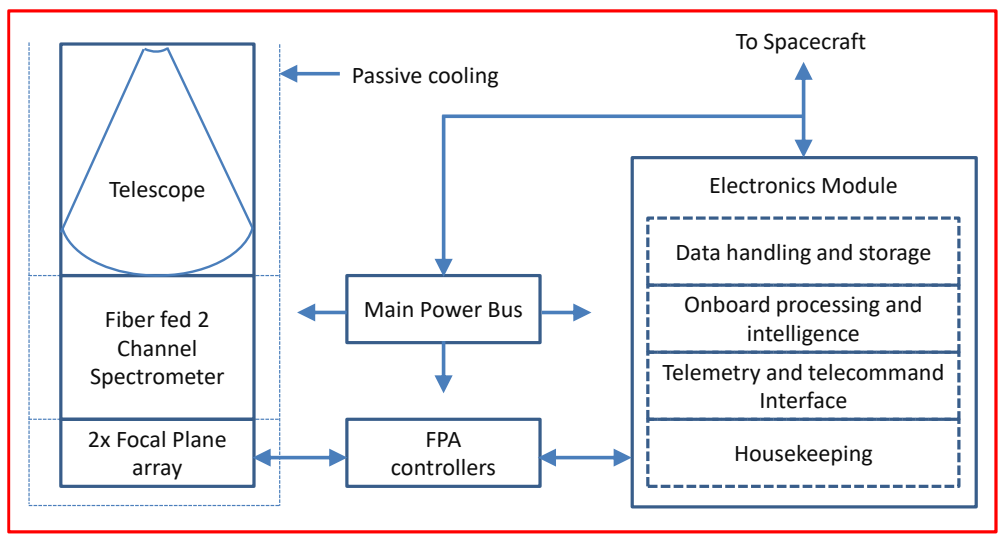

Fig. 2.  IRSIS instrument block diagram

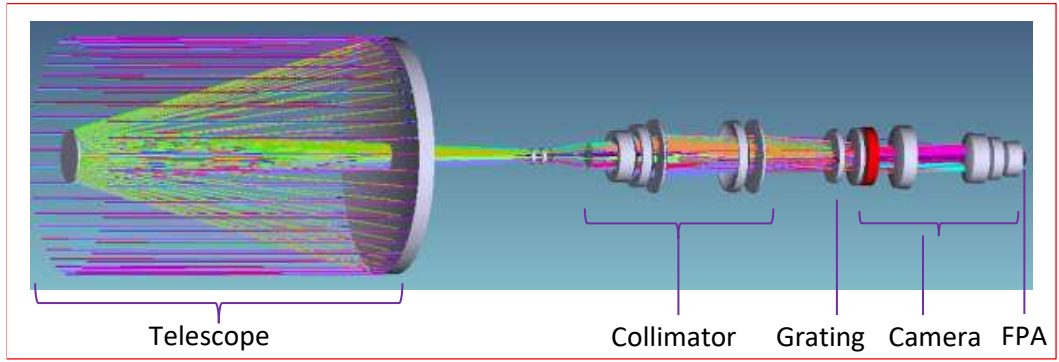

Fig. 3. Zemax ray trace for IRSIS using transmission grating

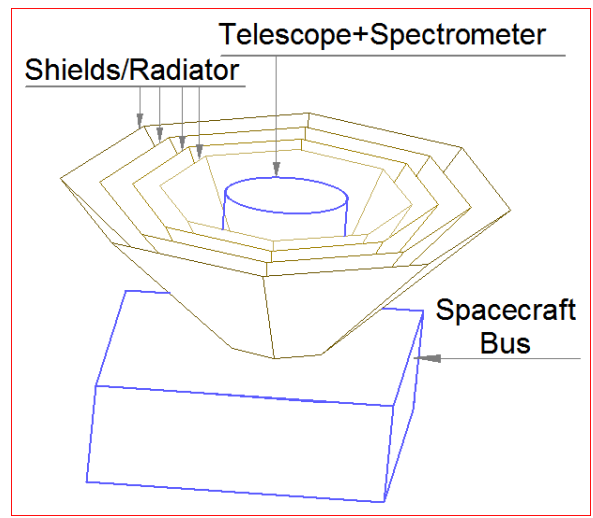

Fig 4. Proposed IRSIS spacecraft bus and V-groove structure

To demonstrate the concept and to gauge the requirements and capabilities to successfully build IRSIS, it was decided to build a laboratory model. The wavelength specifications were limited to 1.7 to 2.5 μm to reduce the costs by using off-the-shelf optical fibres, gratings and lenses and a Functional Grade (2048 x 2048 pixel). Teledyne H2RG mounted on the cold plate of a liquid Nitrogen (LN2) cryostat. A room temperature Teledyne SIDECAR ASIC was used to control the FPA. Since the final aim is to have a simple and lightweight space-based instrument, it was felt that for the IRSIS, we should have a simple FPA controller with not many functions since this will be used as a survey instrument and only few selectable read out modes for the FPAs would suffice for its operation. The SIDECAR is a very complex controller which has been designed for a sophisticated satellite experiment, the James Web Space Telescope (JWST) (Gardner *et al.*, 2006), whereas the IRSIS is a very simple and light payload. We are in the process of building a dedicated FPA controller and data handler for the IRSIS payload which will enable us to have a simple and lightweight controller which can be tailored to our application. Appendix B gives an overview of the laboratory model test results of the IRSIS instrument.

## 3. H2RG Controllers

Following is an overview of a few controllers available.

### 3.1. *SIDECAR ASIC (Teledyne Imaging Sensors)*

SIDECAR ASIC is a product from "Teledyne Imaging Sensors" (Beletic *et al.*, 2008). SIDECAR is an acronym for System for Image Digitization, Enhancement, Control And Retrieval Application. It has been developed for the JWST project. This Application Specific Integrated Circuit (ASIC) which is a semiconductor device, is available in various packages including cryogenic and room temperature versions. Advantages of this product are that it being a product of the H2RG manufacturer, it is closely matched to the H2RG and has small size, low power and programmability. A Jade2 board is interfaced with the SIDECAR ASIC board. The Jade2 board communicates with Windows PC over USB. Teledyne provides Windows compatible software to operate the evaluation board. Teledyne software IDE may be used to compile assembly programs for SIDECAR ASIC. SIDECAR ASIC has 36 Analog channels [16 bit (500 KHz) ADC/12 bit (10 MHz) ADC], 32 Pre amplifiers [Programmable gain 0 to 27 dB in 3 dB steps], 32 programmable digital I/O for clock generation, 20 programmable bias voltage/currents, 16-bit Microcontroller and 32 digital output channel (LVDS or CMOS) digital Interface. Power consumption is less than 100 mW at 100 KHz clocking speed with 32 channel operation. Chip dimensions are about 36 mm x 36 mm.

### 3.2. *ACADIA ASIC*

This is also an ASIC similar to SIDECAR but with lesser complexity developed for the NASA's Wide-Field Infrared Survey Telescope project (Loose *et al.*, 2018). ACADIA is an acronym for ASIC for Control And Digitization of Imagers for Astronomy. Features of ACADIA ASIC are 40 analog channels with programmable gain preamps and 16 bit ADC's, 32 channel clocks with programmable drive, 24 bias channels with programmable current source, MSP430 microcontroller and SPI interface for detector configuration.

### 3.3. *ARC/SDSU/Leach Controller (Astronomical Research Cameras[b])*

This controller is provided by the ARC**.** This is a three-part system, a controller chassis with add-on cards, power supply and Peripheral Component Interconnect (PCI) add-on board for a PC. The vendor provides circuits and source codes for the drivers and application software which are available for Linux and Windows operating systems. Features of the ARC controller include support for many detector types (H2RG arrays and Aladdin II/III arrays, CCDs), 8 Channel IR Video board(s) and bias generator, IR Clock driver board, high speed DSP processor, Fibre optic communication for data, Software and hardware documentation for the controller.

### 3.4. *ESO (European Southern Observatory) NGC Controller*

This ESO New General detector Controller (NGC)[c] has front end basic module and Backplane-Backboard for detector interface. Data is communicated over high speed optic fibre link to PCI add on card at PC. Features of ESO NGC controller are 32 analog channels (32 ADC), FPGA based and high speed optic fibre communication link.

Table 1 shows the detailed specifications of various controllers.

---

[b] www.astro-cam.com

[c] www.eso.org/sci/facilities/develop/detectors/controllers/ngc.html

Table 1. Comparison of FPA controllers

|  | SIDECAR ASIC | ACADIA ASIC | LEACH IR Controller | ESO NGC Controller |
|---|---|---|---|---|
| PROCESSOR | MICROCONTROLLER | MICROCONTROLLER | DSP | FPGA |
| SOFTWARE | ASSEMBLY | ASSEMBLY/C | ASSEMBLY | HDL |
| ANALOG Channels | 32+4 (16/12 bit) | 40 (16 bit) | 8 per board (16 bit) | 32 per board (18/16 bit) |
| DIGITAL Clocks | 32 | 32 | 24 per board | 16 |
| BIAS Voltages | 20 | 24 | 7 per board | 16 |
| Data Communication | LVDS/CMOS | LVDS | Optic fiber | Optic fiber |

## 4. CALIBRATION OF THE H2RG ROIC

In this section we describe the test setup and results of the calibration of the H2RG ROIC.

### 4.1. *Setup*

Fig. 5 shows the SIDECAR ASIC + JADE2 board interfaced with the H2RG ROIC. JADE2 board connects to the PC / Laptop over USB. Teledyne SIDECAR ASIC IDE development software installed in the laptop, communicates with the JADE2 card which in turn communicates with the SIDECAR ASIC. A particular sequence to initialize the SIDECAR ASIC has to be followed. SIDECAR ASIC IDE performs this sequence such as turning on Hardware Abstraction Layer server, loading JADE2 FPGA firmware, powering up SIDECAR ASIC, loading JADE2 registers and uploading SIDECAR ASIC software. These operations are automated if the IDL HxRG GUI provided by Teledyne is used. Further control and image acquisition of SIDECAR ASIC / H2RG were carried out through the IDL GUI. This GUI allows Detector configuration, Preamp configuration, Up-the-ramp sampling configuration and frame capture. Data files are stored in FITS format. For this setup, SIDECAR ASIC is powered through the USB port. Fig. 6A shows the ROIC image captured by the HxRG GUI and Fig. 6B shows a mask (photocopied on a transparency) image captured by H2RG ROIC. These are only for depicting the output.

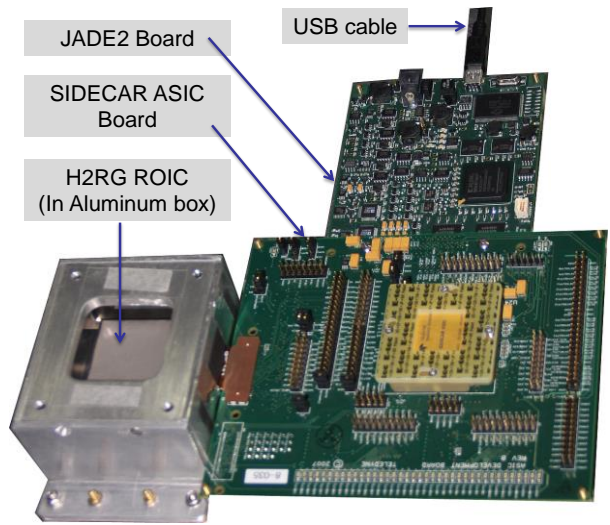

Fig. 5. SIDECAR ASIC and H2RG ROIC setup

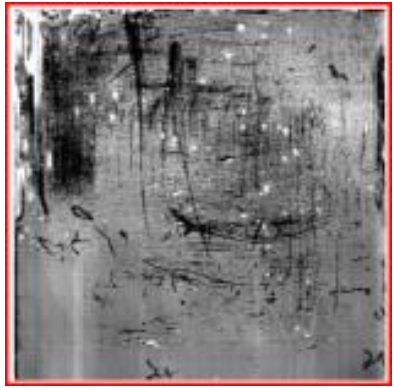 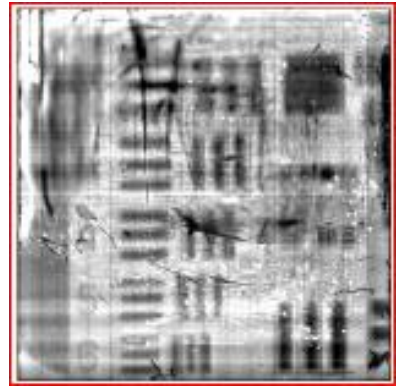

Fig. 6A. H2RG ROIC frame          Fig. 6B. H2RG ROIC frame with mask image

### 4.2. *Conversion Gain*

Conversion gain gives the electrons per ADU. Conversion gain is calculated from SIDECAR programmed settings and later verified by measurement. Preamplifier gain calculation is explained in Table 2.

Table 2. Conversion gain calculation

| |
|---|
| $V_{RP}$ - $V_{RN}$ = (ADC Upper Reference - ADC Lower reference) = 1.691 |
| For Count = 1 and SIDECAR preamplifier set Gain = 1 |
| SIDECAR preamplifier volts/ADU = (1.691/32768) = 51.61 µV/ADU; |
| Transimpedance gain for H2RG (from Teledyne datasheet): 1 electron = 3.55 µV; |
| Hence, Conversion Gain = electrons /ADU = 51.61 µV / 3.55 µV = 14.54 electrons / ADU. |

The SIDECAR gain is measured with SIDECAR input channel 07 fed from battery operated voltage source. Fig. 7 shows the SIDECAR preamplifier gain (slope ~ 19592) (ADU's are referred to as counts in the graphs). From this plot, measured volts/ADU at 0 dB is 51.04 µV and as shown earlier calculated volts/ADU at 0 dB is 51.61 µV.

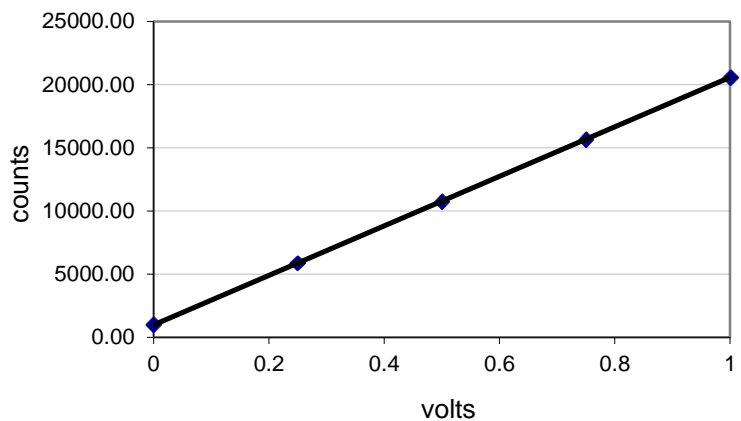

Fig. 7. SIDECAR ASIC Preamp gain plot showing the counts output for different voltage settings

### 4.3. *ROIC Gain*

Fig. 8 shows the graph for source follower gain of ROIC. For this test, the detector bias voltage i.e. reset voltage ($V_{rst}$) is varied and data frames are taken. The counts are converted to voltage. $V_{out}$ is plotted against $V_{reset}$ and the slope gives the source follower gain. The gain measured is about 0.81.

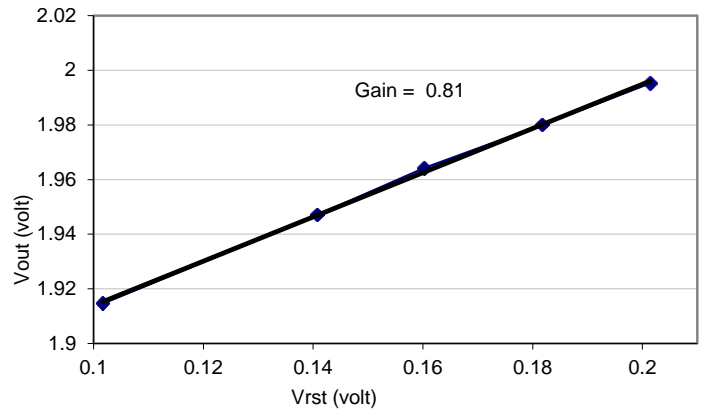

Fig. 8. H2RG ROIC gain plot showing Vout *versus* Vrst

## 5. CALIBRATION OF THE H2RG FPA

In this section we discuss the calibration setup and test results of the calibration of the H2RG FPA.

### 5.1. *Setup*

A setup for carrying out various FPA tests is shown in Fig. 9. This setup shows the LN2 cryostat with a functional grade H2RG FPA, SIDECAR-JADE2 controller board, Lakeshore temperature controller, blackbody source and data acquisition laptop. The H2RG needs to be cooled to about 80 Kelvin at a rate slower than 2 Kelvin per minute. This was achieved by controlled pouring of LN2 into the cryostat and setting up the Lakeshore Temperature Controller ramped temperature control option to 1.5 Kelvin per minute. Fig. 10 shows the cooling curves for H2RG for FPA along with other locations in the setup.

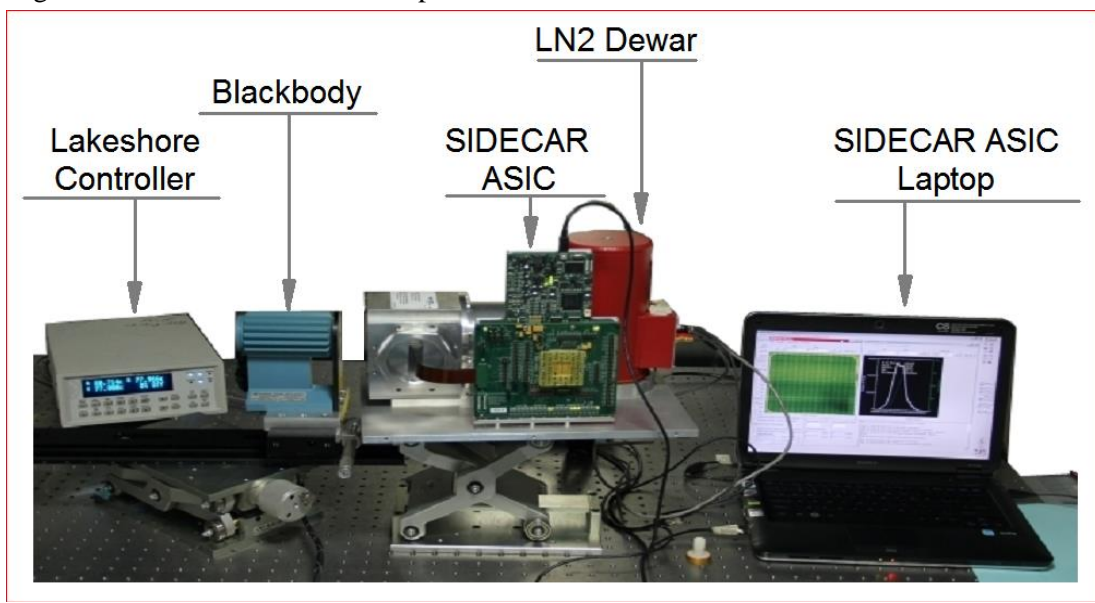

Fig. 9. H2RG FPA test setup

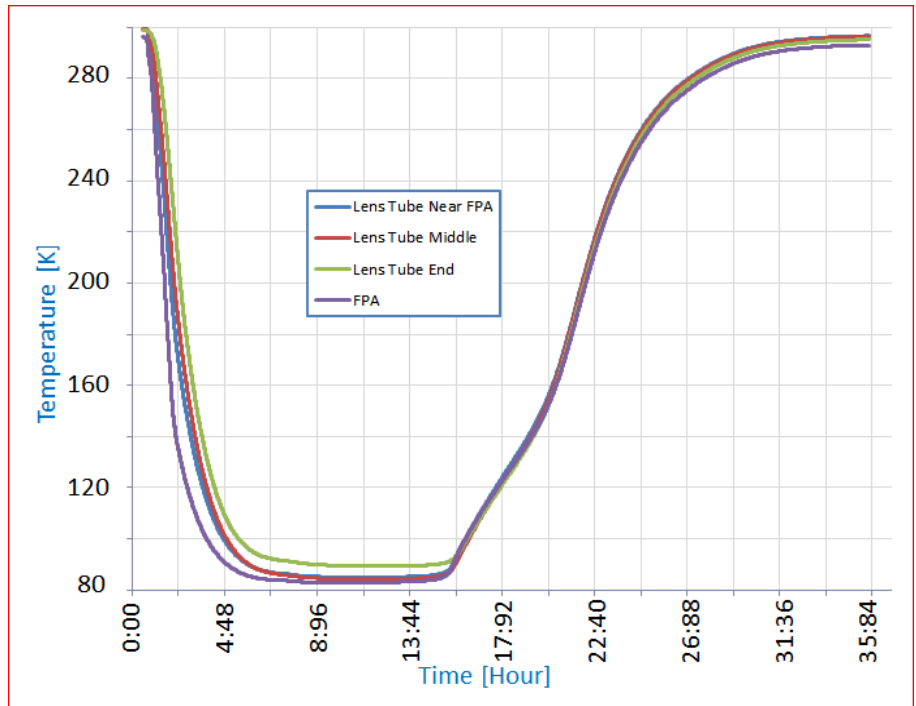

Fig. 10. Cooling and warmup cuves for H2RG FPA setup

### 5.2. Well capacity

Well capacity measured is the maximum number of electrons that can be collected by a photo diode pixel (at applied bias). Well capacity was measured by two sets of measurements. For one set of data, FPA was exposed to photons from a black body at 393 Kelvin (BB393K) for near saturation counts. For another set of data, FPA was exposed to photons from 80 Kelvin (Blank) aluminum disk on the filter. Both data sets were taken in the up-the-ramp mode. The mean of "Blank" data set was 41700 whereas the mean of BB393K was 49900, yielding a difference of 8200 counts, which results in 119228 electrons as well capacity of a pixel at 256 mV bias.

### 5.3. Dark current

Dark current is a measure of the signal generated by the FPA pixels when no light is falling on it and is dependent on the FPA temperature. For this test, the FPA was exposed to a blank made up of an aluminum cap at 80 Kelvin. Data was taken in the up-the-ramp mode for about 200 seconds. The rate of counts is converted to electrons. Dark current was measured to be < 1 electron/sec.

### 5.4. Read noise

For this test, the same setup as for the dark current test was used and data was taken in the up-the-ramp mode. The difference of 10 consecutive frames was used to calculate the read noise. The median read noise value was estimated to be about 100 electrons. The higher read noise is probably due to functional grade nature of H2RG FPA, **room temperature** version of SIDECAR kit, 15 inch long flex cable, preamplifier configuration related noise and possible power supply voltage ripple (see e.g., Wong *et al.,* 2004; Loose *et al.*, 2005; Loose, 2007).

### 5.5. Quantum Efficiency

The setup for quantum efficiency (QE) test is shown in Fig. 11. The black body source is placed at a distance of 30 mm from the dewar window. Inside the dewar, an aluminum cap with 3 mm hole at the centre is placed over the cold filter to reduce the amount of radiation reaching the detector. The data for various temperatures of the black body source was taken in the up-the-ramp mode. The ray trace model of the setup using ZEMAX software is shown in Fig. 12. This model gives the number of photons reaching the detector. The H2RG detector has 8 rows and 8 columns of reference pixels. Using these reference pixels, correction in the offset of image pixels is performed. Figs. 13A and 13B show the comparison of the unprocessed and the reference row/column processed images for the blackbody at 323 Kelvin. Fig. 14 shows the response of the FPA at various blackbody temperature settings.

QE calculations are done as follows. Total photons radiated by the blackbody at a temperature T are calculated for wavelengths between 1.5 to 2.5 µm. This number is fed to ZEMAX model of the camera. Details of window and filter material are also fed to the ZEMAX model. ZEMAX simulation yields the number of photons per second per pixel to FPA. From the dataset for blackbody temperature T, counts per seconds per pixel are calculated. These are then converted to electrons per second per pixel (1 count = 14.54 electrons). QE is calculated as the ratio of electrons/sec/pixel to photons/sec/pixel. For reference row column processed case, QE calculated was ~ 40 % (for 1.5 to 2.5 µm) and for reference row column unprocessed case, QE calculated was ~ 44 % (for 1.5 to 2.5 µm).

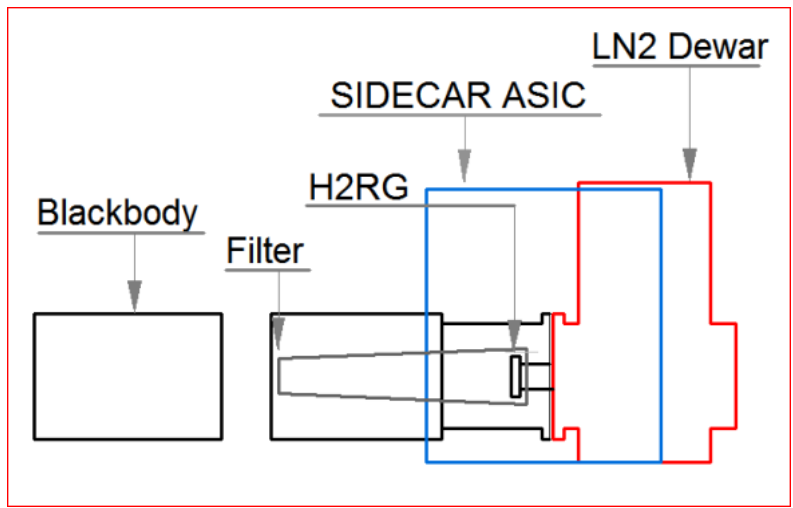

Fig. 11. H2RG FPA QE test setup

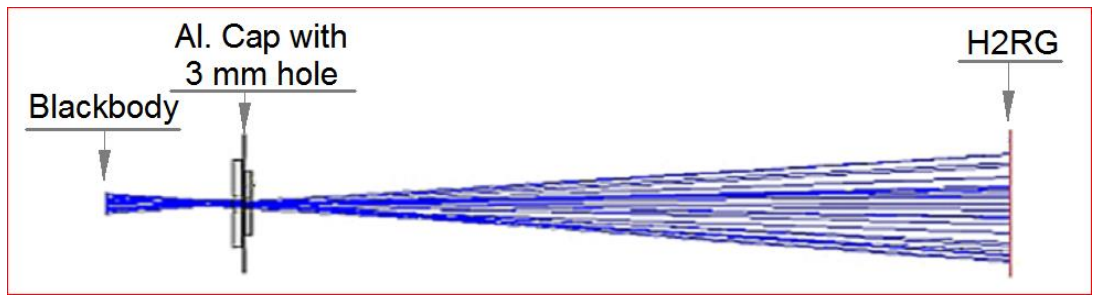

Fig. 12: ZEMAX ray trace model for QE setup

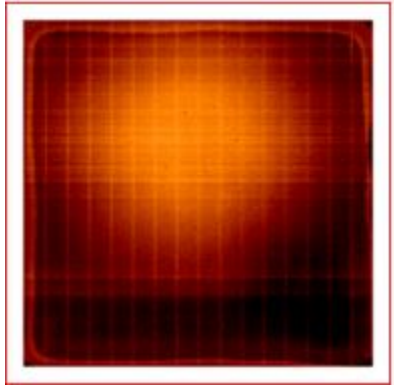

Fig. 13A. H2RG FPA raw image for blackbody

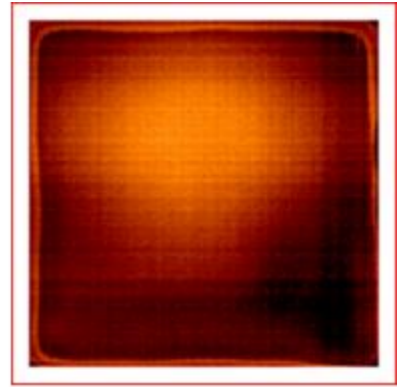

Fig. 13B. Reference row/column processed H2RG FPA image for blackbody

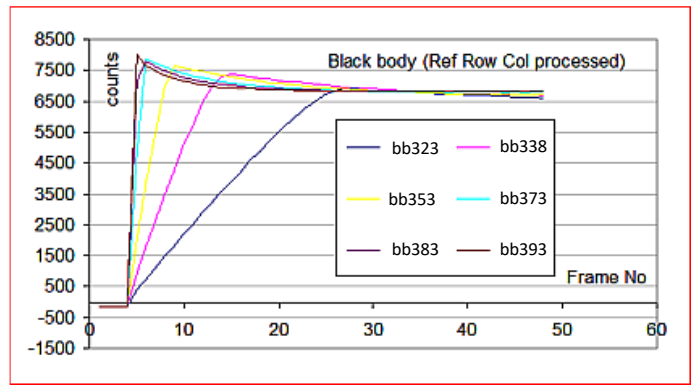

Fig. 14. H2RG FPA response plot (counts *versus* frame numbers) at various blackbody temperatures (in Kelvin)

## 6. The TIFR IR Group prototype controller

As a dedicated FPA controller is required for the IRSIS satellite project, a fully customized H1RG/H2RG controller is being developed in-house. A prototype using a microcontroller has been designed and tested with the H2RG ROIC.

### 6.1. *Prototype FPA Controller Setup*

The block diagram of the prototype is shown in Fig. 15. All the clocks required for the H2RG ROIC (Teledyne H2RG datasheet) are generated by the microcontroller. DAC's are programmed to generate power and bias voltages required for the H2RG ROIC. Analog output signals from the H2RG ROIC are pre-amplified by an Op-amp buffer and digitized by a 10-bit ADC within the microcontroller. Data is sent to the computer over the USB port. The prototype TIFR IR Group FPA controller is shown in Fig. 16. The H2RG ROIC along with its mounting frame is installed in a milled aluminum box. A PGA to HIROSE connector flex cable is connected from the ROIC to the fan-out board. The fan-out board (see Fig. 17) is a two layer PCB with a provision of two sets of header connectors and two 50 pin D connectors. The fan-out board interfaces with the controller board via jumper cables. The major components of the FPA controller are an Advance RISC Machine (ARM) microcontroller, DAC Bias generator with driver and Pre-amplifier. The DAC has 12 bit input and can be programmed over the serial bus. The DAC can be programmed for output voltages from 0 to 3.3V. The DAC output is buffered by an operational amplifier buffer. The Clock buffer/driver stage is optional with this setup. A source follower drive may be used to drive the internal source follower transistor of the H2RG ROIC. In the current configuration, a resistor from $V_{dd}$ bias serves as a source follower drive. A unity-gain operational amplifier acts as the pre-amplifier stage. All antistatic precautions have to be taken while handling and connecting the H2RG ROIC to avoid damage. Prior to applying power and biases, they were measured and applied in the correct sequence. At POWER ON, the clocks generated by the microcontroller were set to their inactive states.

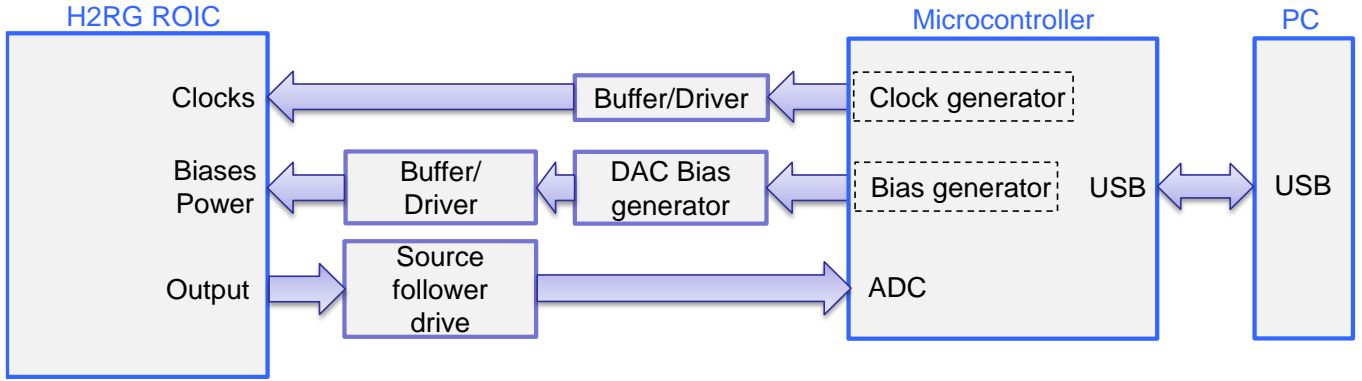

Fig. 15. Block diagram of H2RG ROIC prototype controller

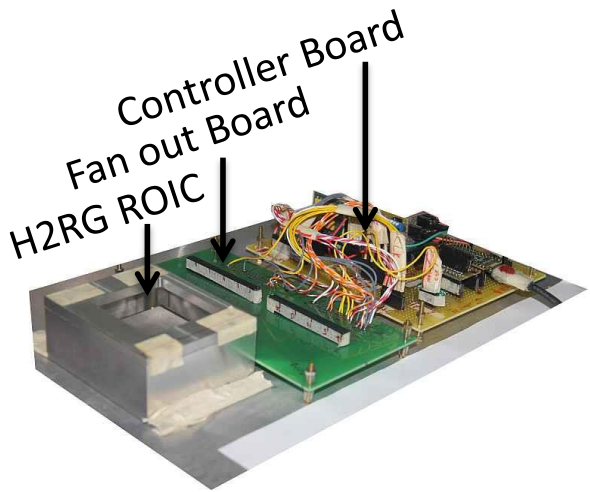

Fig. 16. Prototype of TIFR IR Group H2RG ROIC controller

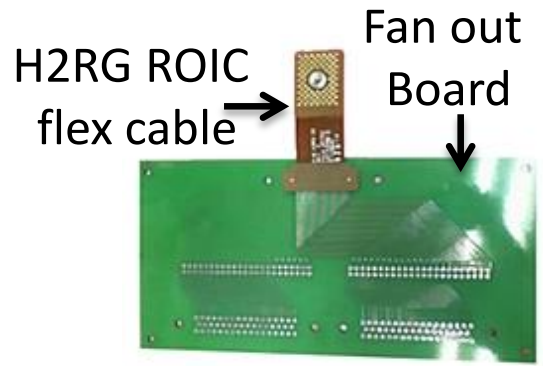

Fig. 17 Fan-out board with H2RG ROIC flex cable

### 6.2. *Power and bias generation*

The DAC's are programmed to generate power and biases required by the H2RG ROIC. Fig. 18 shows the clocking signals generated by the microcontroller to program one of the DACs. The DAC internal register is programmed by CLK and SDI (serial data in) and latched with the LOAD pulse. Fig. 19 shows the response of the DAC for various programmed values.

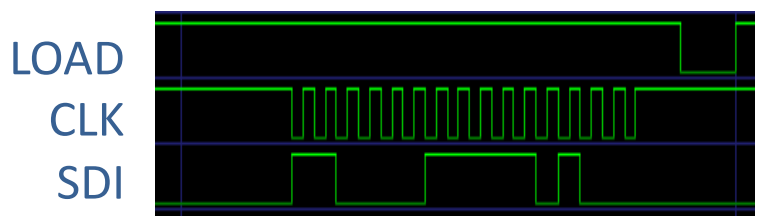

Fig. 18. DAC programming clocks for generating H2RG ROIC bias voltages

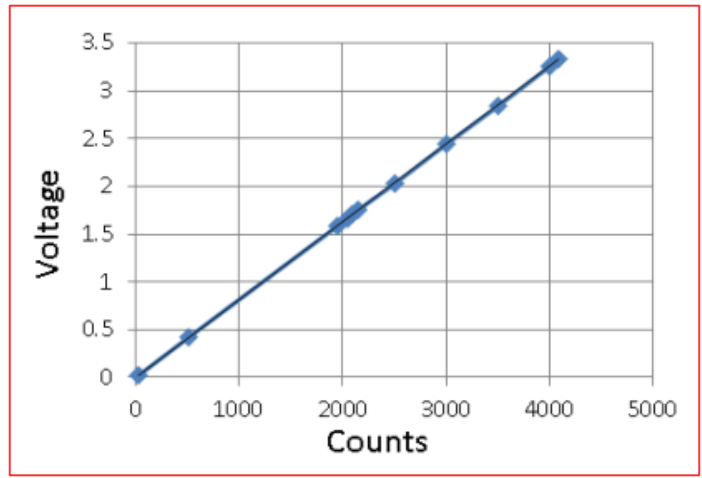

Fig. 19. DAC test for generation of bias voltage (output voltage *versus* programmed value)

### 6.3. *Clock generation*

Fig. 20 shows the pattern generated by the microcontroller. This pattern is just the demonstration of the real clock pattern required to operate the FPA. The number of clock pulses shown in the figure are much less since during the actual H2RG ROIC clocking cycle 2048+ clock pulses will be required per line of readout for a single channel readout.

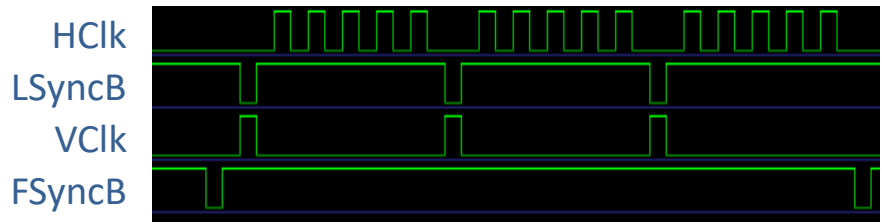

Fig. 20. H2RG FPA clocks generated by the microcontroller

### 6.4. *H2RG initialization*

The H2RG has several programmable registers. These registers provide different functionality such as the number of outputs to be used, output buffer selection, window mode and window mode coordinate selection. Fig. 21 shows the clocks generated by the microcontroller for programming the H2RG ROIC internal registers through its serial programming interface. DATACLK and DATAIN lines can be shared with the VCLK and FSYNCB signals of the vertical scanner. In this prototype, VCLK is used in place of DATACLK and FSYNCB is used in place of DATAIN. The CSB signal enables serial interface block for communication. Two ROIC registers were programmed. One to enable buffered output and another for single output mode.

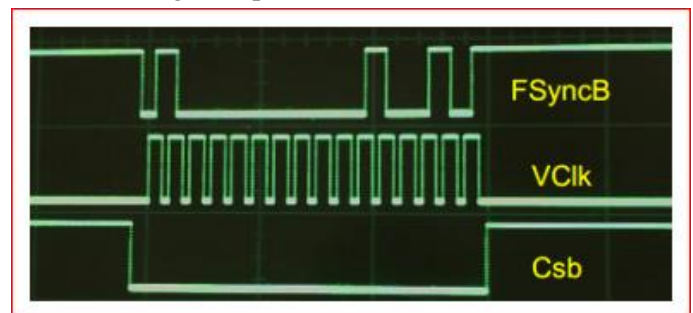

Fig. 21. Clocks to program internal registers of H2RG ROIC

### 6.5. *H2RG data capture*

The H2RG was programmed for single output mode and the frame clocking was started. The analog signal output from the ROIC is shown in Fig. 22. This signal is digitized by the ADC within the microcontroller.

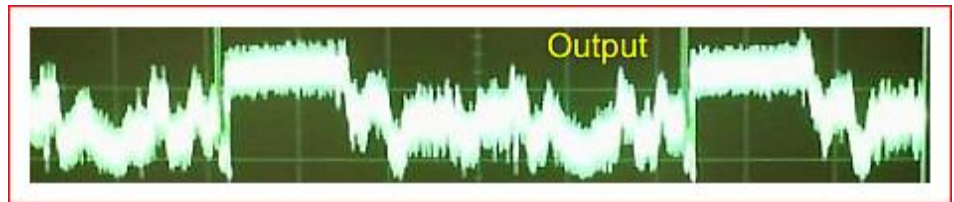

Fig. 22. H2RG ROIC output

### 6.6. *Software flowchart*

The software consists of two modules: one residing in the ARM microcontroller and the other in the PC. These control the overall operation of the controller. Fig. 23A is a flowchart showing the C++ program flow at the microcontroller and Fig. 23B shows the flowchart for the PC program. For the microcontroller, all the clocks are kept in their inactive states. Power and biases are generated by programming the DAC and are applied to the ROIC. The ROIC internal registers are programmed and clocking is started for a frame after the user request from the PC. At the PC, data is captured over the USB port and stored in a buffer. On receipt of the complete frame, the data is written to a FITS file.

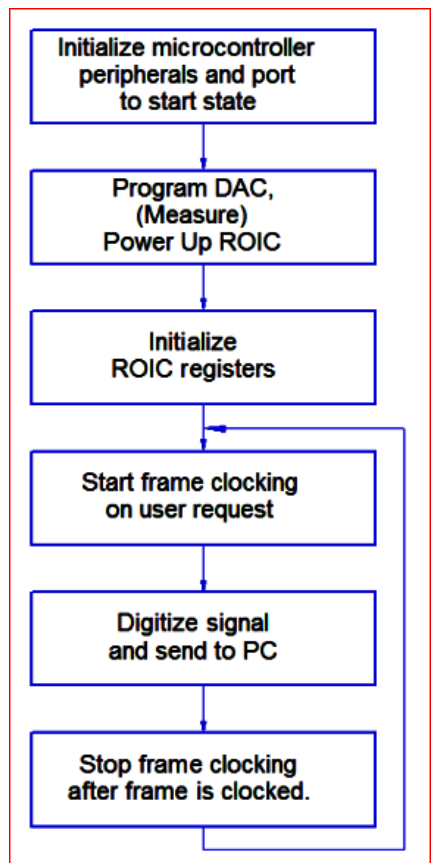 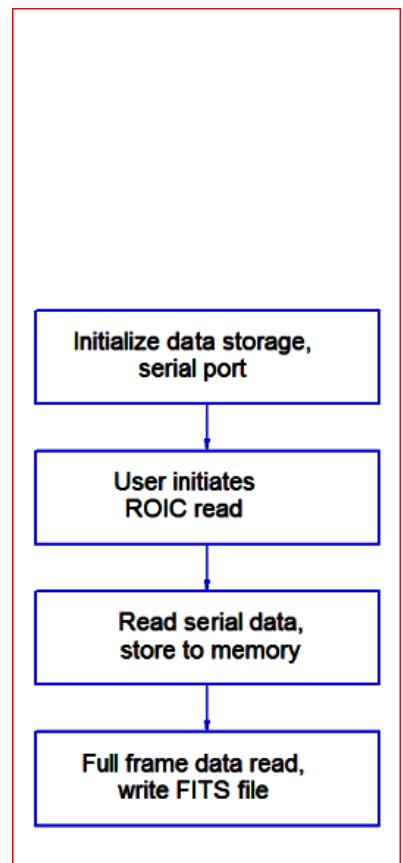

Fig. 23A. Microcontroller program flowchart     Fig. 23B. PC program flowchart

### 6.7. *Sample of captured frame*

Fig. 24A shows the H2RG ROIC frame capture done with the SIDECAR ASIC controller and Fig. 24B shows the frame capture done with the TIFR IR Group prototype controller. Fig. 25 is an image of a mask on the ROIC. These images are only for depicting the output of the ROIC using different controllers since the ROIC is not optimized for imaging.

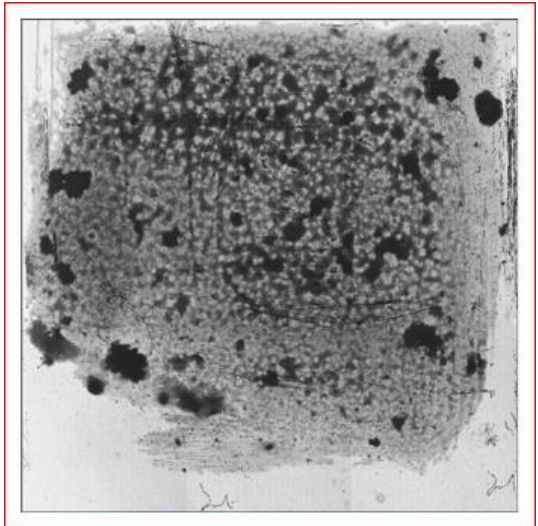
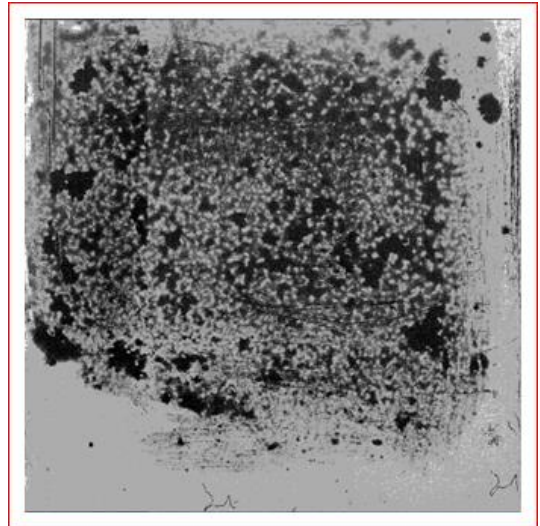

Fig. 24A. H2RG ROIC capture with the SIDECAR ASIC

Fig. 24B. H2RG ROIC capture with the TIFR IR Group prototype controller

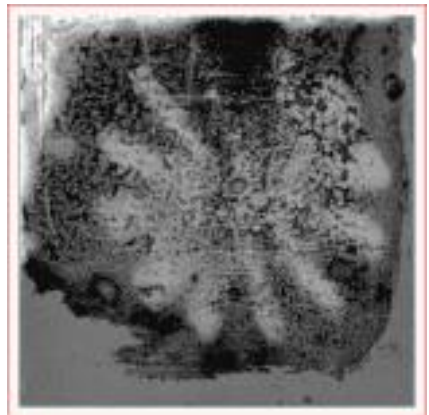

Fig. 25. H2RG ROIC capture with mask with TIFR IR Group prototype controller

### 7. Summary and future work

A simple prototype controller was built to prove the ability of the in-house controller to successfully clock the H2RG detector and acquire and store images as per the requirements of the space-based spectrometer payload. Unfortunately, the ROIC used was old and degraded and is anyway not designed to obtain images since the image section is missing in an ROIC and at most the intrinsic sensitivity of Silicon to visible light can be used to obtain average quality images.

After gaining confidence with the simple controller, a more professional controller design is now under development using an FPGA as a primary control element. Fig. 26 shows the proposed H1RG controller block diagram. The FPGA will handle functionalities like initialization of various bias and power supplies to the FPA and preamplifier section via a DAC section, initialize the H1RG FPA via a separate serial interface and generate clocks for the FPA. The FPGA will also control the digitization of output data from the FPA and stream the data to memory banks. A preprocessor will be used for first-order data processing of the data available in the memory banks. A microcontroller will be used in supervisory mode to handle commands and other miscellaneous functions of the controller. The choice of components will be primarily from the ISRO inventory of flight tested devices and components with heritage would be used. Bias voltage generation via DAC and FPGA has already been tested and FPA clocks have been generated using an FPGA. This proposed in-house controller will be used in the IRSIS flight model.

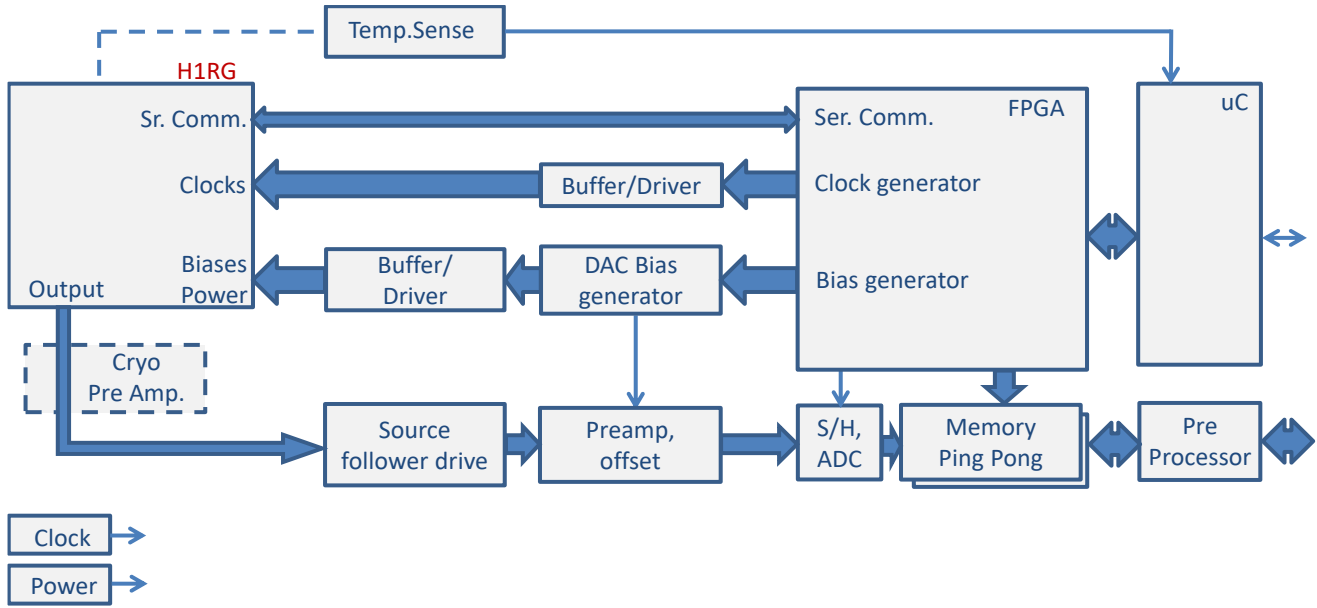

Fig. 26. Proposed H1RG controller block diagram

**Acknowledgments**

We thank the anonymous reviewer for several useful comments and suggestions, which greatly improved the presentation of the paper. Support and help of all the IR group and DAA members are highly appreciated. We acknowledge the support of the Department of Atomic Energy, Government of India, under project identification no. RTI 4002.

# **Appendices**

## **Appendix A: H2RG ROIC and FPA**

Fig. A1 shows an H2RG ROIC with flex cable. The H2RG has 2048 x 2048 pixels of which there are 8 rows (4 at the top and 4 at the bottom) and 8 columns (4 to the left and 4 to the right) of reference pixels. Pixel pitch is 18 µm and the FPA's overall size is about 39 mm x 41 mm x 21 mm (L x W x H). The following numbers are the typical rating for the H2RG from the Teledyne data sheets and can vary for each piece. Optimum operating temperature is 37 Kelvin, median read noise is <18 electron rms and median dark current <0.05 electrons/sec. The well capacity is about 123000 electrons. The FPA can be operated in imaging and/or guide mode and has 32 outputs with a choice of using either 32, 4 or 1 output. The output mode could be un-buffered or buffered. The FPA can be read out at 100 KHz or at 5 MHz in a special fast mode. At 100 KHz readout speed, power dissipation is about 800 microwatts/channel for the buffered mode. The H2RG FPA is made up of several blocks. The detector layer contains photo sensitive pixels and the reference pixels are passive capacitor arrays arranged in 4 rows and 4 columns at the edges of the array. The serial interface block allows communication with the FPA for programming the internal registers. Slow and fast scanners (decoders and shift register) address pixels for full frame array/sub-frame array readout. A separate fast column buffer readout circuitry is provided for 5 MHz readout speed. There are two major sections of readout circuits responsible for transferring electrons from the photo diode pixel through the indium bump interconnect to the output pad of the FPA.

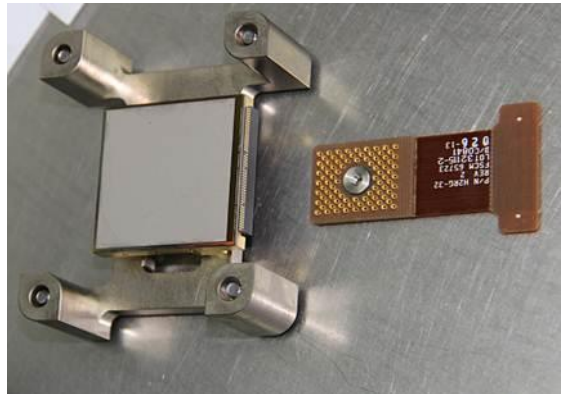

Fig. A1. H2RG ROIC and flex cable

## **Appendix B: Overview of the IRSIS Laboratory Model Test Results**

### **B.1 Laboratory Model tests**

Fig. B1A shows the model proposed for the camera module and the H2RG FPA interface with LN2 dewar. Though Stirling coolers were tested for performance, considering the need to cool optical component also, LN2 dewar model was chosen. These were implemented as shown in Fig. B2A and Fig. B2B.

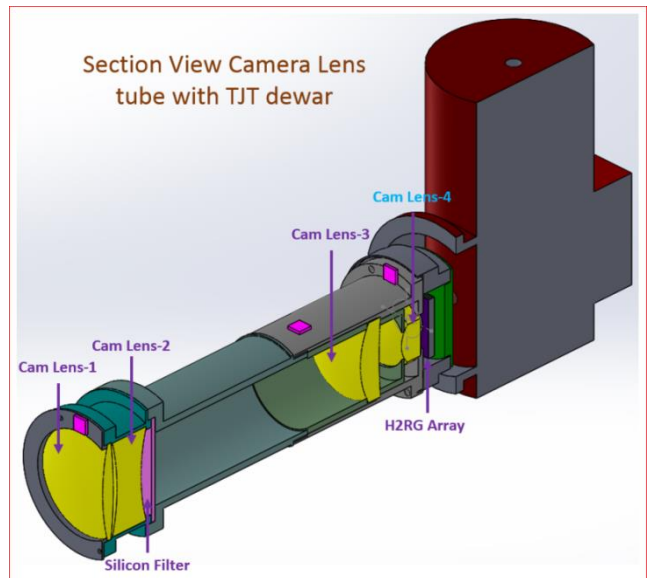

Fig. B1A. H2RG FPA – camera interface (LN2 dewar version)

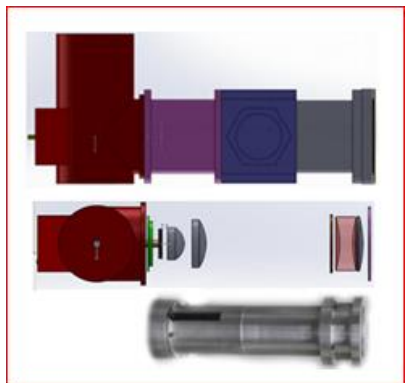

Fig. B2A. Schematic of H2RG and camera interface with LN2 dewar

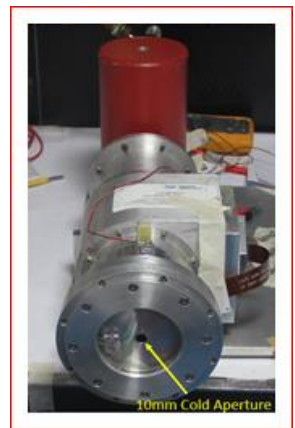

Fig. B2B. Completed H2RG and camera interface with LN2 dewar

## *B.2 Lab model results*

Fig. B3 shows the schematic of the laboratory model test setup of the shortwave spectrometer. Blackbody or a Xenon gas tube is used as the light source for this setup. Light collected by the focal plane fibre plate is transferred to the slit fibre plate where the slit is formed. The slit fibre plate is coupled to the Collimator + Grating module which in turn is coupled to the Camera + H2RG section in the LN2 dewar. Fig. B4 shows the focal plane fibre plate and the slit plane fibre plate connected with optical fibres. Fig. B5A shows the focal plane fibre plate and Fig. B5B shows the slit plane fibre plate, with the fibres illuminated with light. A photograph of this setup is shown in Fig. B6 with captions of various parts. Fig. B7 shows the comparison of spectra obtained with the above setup for blackbody at 773 Kelvin and ZEMAX simulation.

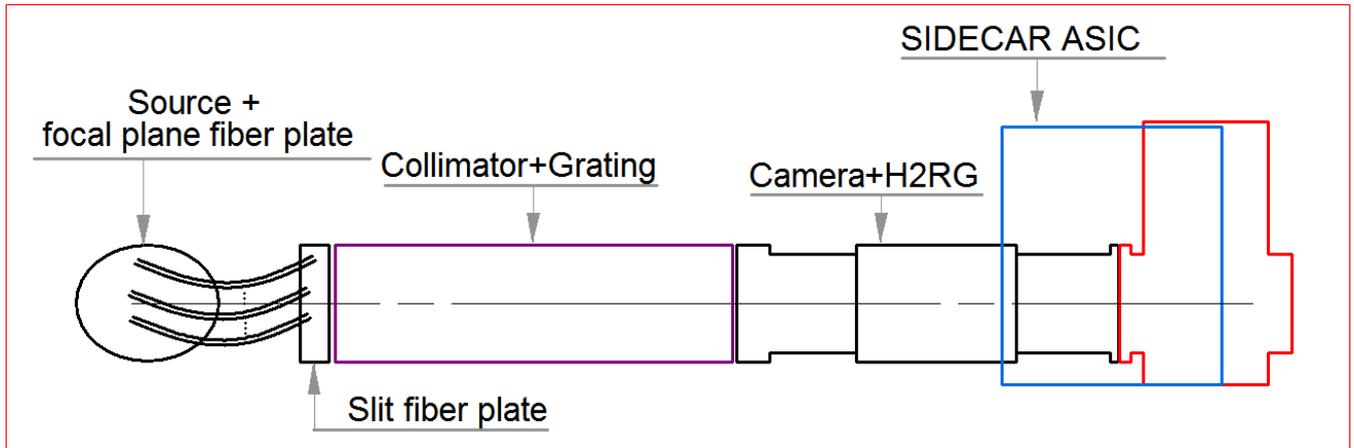

Fig. B3. Test setup schematic for the shortwave spectrometer (laboratory model)

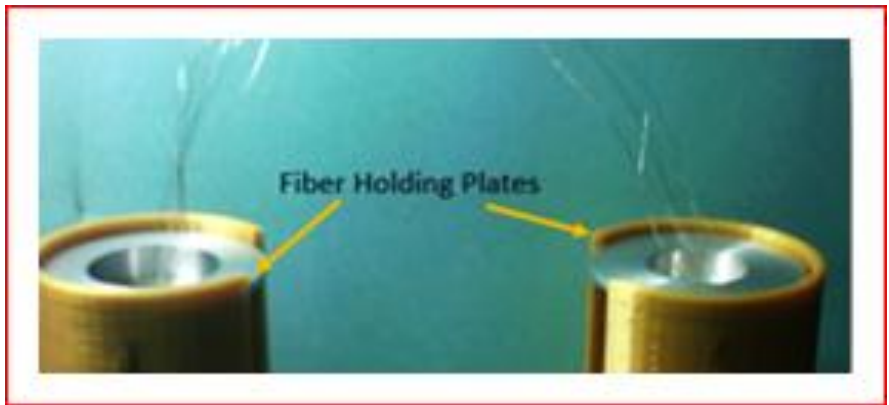

Fig. B4. Focal plane fiber plate and slit plane fiber plate connected with optic fibers

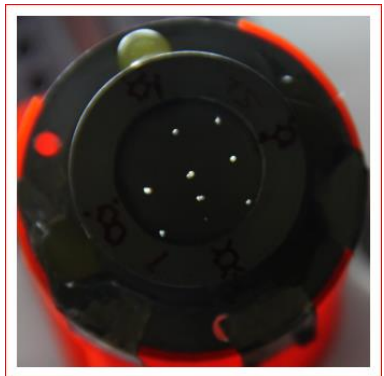 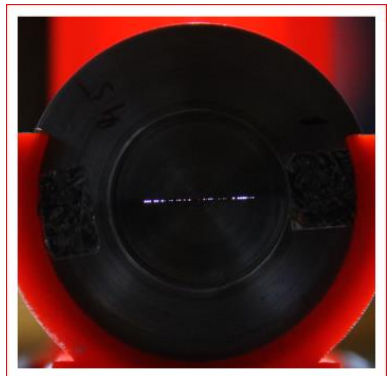

Fig. B5A. Focal plane fiber plate     Fig. B5B. Slit plane fiber plate

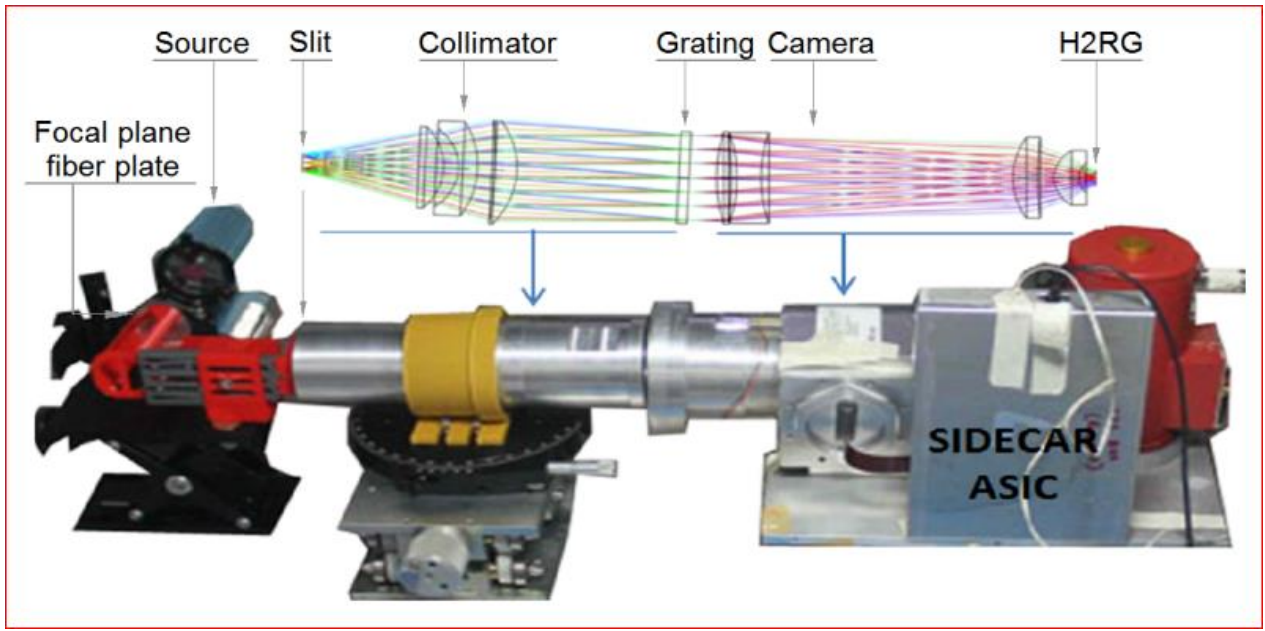

Fig. B6. Test setup for the Shortwave spectrometer (laboratory model)

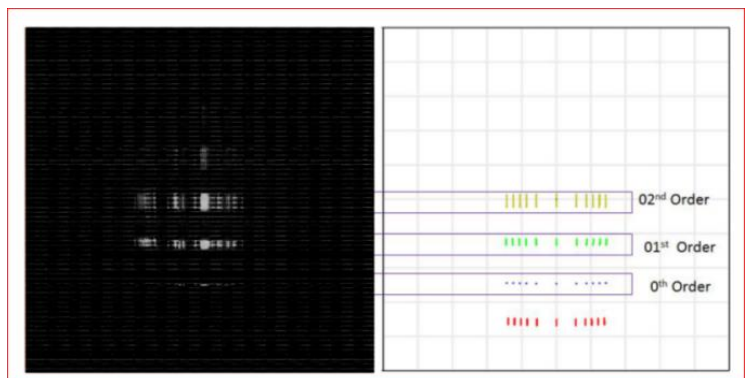

Fig. B7. Comparion of spectra obtained from test setup (left) with Zemax modelling (right)

**References**

Beletic, J. W., Blank, R., Gulbransen, D., *et al.* [2008] *Proc. of SPIE* 7021, 70210H.
Bolton, D. A. [2017] *47th International Conference on Environmental System*s, ICES-2017-360, 16-20 July 2017, Charleston, South Carolina.
Cornut, M., Riti, J.-B., Hauser, A., Rauscher, U. & Collaudin, B. [2000] *SAE Transactions* 109, 456, Published by: SAE International.
Gardner, J. P., Mather, J. C., Clampin, M., *et al.* [2006] *Space Sci. Rev.* 123, 485.
Ghosh, S. K. [2010] ASI Conference Series 1, 171, Edited by D. K. Ojha.
Hasebe, T., Kashima, S., Uozumi, S., *et al.* [2018] *Proc. of SPIE* 10698, 1069864.
Loose, M., Beletic, J., Blackwell, J., *et al.* [2005] *Proc. of SPIE* 5904, 59040V.
Loose, M., Beletic, J., Garnett, J. & Xu, M. [2007] *Proc. of SPIE* 6690, 66900C.
Loose, M., Smith, B., Alkire, G., *et al.* [2018] *Proc. of SPIE* 10709, 107090T.
Wong, S. S., Loose, M., Piquette, E. C., Garnett, J. D., Zandian, M., & Farris, M. C. [2004] *Proc. of SPIE* 5499, 258.